\documentclass[trackchanges,twocolumn]{aastex7}

\usepackage{comment}
\usepackage{hyperref}
\usepackage{braket}
\usepackage{amsmath}

\newcommand{\modify}[1]{  #1}

\begin{document}

\title{Confined Circumstellar Material as a Dust Formation Site in Type II Supernovae}

\author[orcid=0000-0002-8215-5019,gname=Yuki, sname='Takei']{Yuki Takei}
\affiliation{Yukawa Institute for Theoretical Physics (YITP), Kyoto University, Kitashirakawa-oiwake-cho, Kyoto, Kyoto 606-8502, Japan}
\affiliation{Research Center for the Early Universe (RESCEU), School of Science, The Unviersity of Tokyo, 7-3-1 Hongo, Bunkyo-ku, Tokyo 113-0033, Japan}
\affiliation{Astrophysical Big Bang Laboratory, RIKEN, 2-1 Hirosawa, Wako, Saitama 351-0198, Japan}
\email[show]{yuki.takei@yukawa.kyoto-u.ac.jp}

\author[orcid=0000-0002-3517-1956,gname=Kunihito, sname='Ioka']{Kunihito Ioka} 
\affiliation{Yukawa Institute for Theoretical Physics (YITP), Kyoto University, Kitashirakawa-oiwake-cho, Kyoto, Kyoto 606-8502, Japan}
\email{kunihito.ioka@yukawa.kyoto-u.ac.jp}

\author[orcid=0000-0002-4979-5671,gname=Masaru, sname='Shibata']{Masaru Shibata} 
\affiliation{Max Planck Institute for Gravitational Physics (Albert Einstein Institute), Am M\"uhlenberg, 1, Postdam-Golm 14476, Germany}
\affiliation{Yukawa Institute for Theoretical Physics (YITP), Kyoto University, Kitashirakawa-oiwake-cho, Kyoto, Kyoto 606-8502, Japan}
\email{masaru.shibata@aei.mpg.de}

\begin{abstract}
We propose a model for dust formation in Type II supernovae (SNe) interacting with confined circumstellar material (CSM), motivated by recent time-domain surveys that have revealed a substantial fraction of SN progenitors to be surrounded by CSM ejected shortly before core-collapse.
We simulate the pre-SN mass eruption and the resulting confined CSM using the open-source code \texttt{CHIPS}, and follow the subsequent evolution of the SN ejecta and its interaction with the CSM.
We show that a cold dense shell (CDS) is formed at the radiative shock under a wide range of conditions and later undergoes rapid adiabatic cooling during free expansion, leading to efficient dust condensation.
The resulting dust mass ranges from $\sim10^{-3}\,M_\odot$ to $0.1\,M_\odot$, depending on the mass and spatial extent of the CSM.
We further calculate the infrared (IR) emission from the newly formed dust and find broad consistency with observations of SN~1998S.
Notably, the IR light curve exhibits a rapid rise within $\lesssim10\,{\rm d}$, closely resembling that of kilonovae (KNe).
This suggests that dust emission powered by confined CSM interaction may be also discovered in KN searches.
Moreover, the high-density environment of the CDS may allow dust grains to grow to larger sizes, enhancing their survivability against destruction by reverse shocks propagating from the interstellar medium at later times.
\end{abstract}


\section{Introduction} 
\label{sec:Introduction}
Core-collapse supernovae (CCSNe) are among the most energetic events in the universe and play a key role in shaping the chemical and dynamical evolution of galaxies \citep[e.g.,][]{Woosley_Weaver_1995,Nomoto_et_al_2006,Tsuna_et_al_2023}. These explosions not only release large amounts of energy and nucleosynthetic products into the interstellar medium (ISM) but are also considered a significant source of cosmic dust, especially in the early universe \citep[e.g.,][]{Todini_Ferrara_2001,Nozawa_et_al_2003,Bianchi_Schneider_2007,Dwek_et_al_2007}.

SNe IIn are a subclass characterized by narrow hydrogen emission lines, indicating strong interaction between the SN ejecta and a dense circumstellar material (CSM) that had been expelled shortly before the explosion \citep{Schlegel_1990}. In these events, efficient radiative cooling of the shocked gas leads to the formation of a dense shell, which provides favorable conditions for dust formation \citep[e.g.,][]{Pozzo_et_al_2004,Mattila_et_al_2008,Tinyanont_et_al_2016}. Indeed, infrared (IR) observations of SNe IIn such as SN 2005ip and SN 2010jl have shown clear signatures of newly formed dust \citep[e.g.,][]{Fox_et_al_2010,Maeda_et_al_2013,Gall_et_al_2014}, highlighting the importance of CSM interaction as a dust production mechanism.

More recently, observational studies have uncovered compact and dense CSM structures, often referred to as ``confined CSM", in a broader population of SNe II, with typical radii of the confined CSM being $\lesssim10^{15}\,{\rm cm}$ \citep[e.g.,][]{Morozova_et_al_2017,Forster_et_al_2018,Bruch_et_al_2021,Hiramatsu_et_al_2023,Dastidar_et_al_2024}. While the mechanisms responsible for driving such mass loss remain poorly understood, these structures are thought to originate from late-stage eruptive, wave-driven mass loss during the final years of stellar evolution \citep[e.g.,][]{Quataert_Shiode_2012,Fuller_2017,Wu_Fuller_2022a}, or a binary interaction \citep{Chevalier_2012,Metzger_2022,Wu_Fuller_2022}. Their presence is also supported by early-time light curves \citep{Morozova_et_al_2018} and flash spectroscopy \citep{Yaron_et_al_2017}. This growing body of evidence suggests that interaction-powered transients and dust formation in dense CSM environments may be more ubiquitous among CCSNe than previously recognized \citep[$\gtrsim36$\% of all SNe II progenitors,][]{Bruch_et_al_2023,Hinds_et_al_2025}.

Dust formation in SNe requires that the gas reaches sufficiently low temperatures and high densities for condensation to occur \citep[e.g.,][]{Kozasa_et_al_1989,Nozawa_et_al_2003,Nozawa_et_al_2010}. Such conditions are naturally established in cold dense shells (CDSs) produced by radiative shocks, particularly when a confined and dense CSM is present. The shock heats the gas to post-shock temperatures of $\gtrsim10^{7}\,{\rm K}$, after which the gas cools efficiently via radiation, mainly through free-free emission. The high density associated with confined CSM enhances the cooling efficiency, allowing the temperature to drop to $\sim10^{4}\,{\rm K}$ immediately. This leads to the formation of a geometrically thin shell, which provides favorable conditions for dust condensation. Therefore, the evolution of the CDS is a key factor in determining the amount of dust that can form in an SN event.

One of the few theoretical studies that have investigated dust formation in SNe interacting with the dense CSM is that by \citet{Sarangi_Slavin_2022}, who examined dust production in SNe IIn with extended CSM formed by steady stellar wind. Their work explored the conditions under which dust can form in the CDS and discussed the resulting properties of dust grains. While their study has provided important insight into dust formation in such environments, it focused on CSM structures that are significantly more extended and less dense than the confined CSM inferred from recent early-time observations of SNe II. The physical conditions associated with confined CSM remain less explored in the context of dust formation, despite their increasing observations.

In this work, we investigate the conditions for CDS formation and dust production in a confined CSM environment. We compute the pre-explosion mass-loss history using the open-source code Complete History of Interaction-Powered Supernovae (\texttt{CHIPS})\footnote{\texttt{CHIPS} and documentation including usage instructions for the code is available in the following link: \url{https://github.com/DTsuna/CHIPS.git}} \citep{Takei_et_al_2022,Takei_et_al_2024} to model the formation of confined CSM, and then calculate the post-explosion shock evolution using the thin-shell approximation. By estimating the mass of the CDS formed in these interactions, we aim to clarify how the structure of confined CSM influences the efficiency of dust formation in SNe II.

This paper is constructed as follows: In Section \ref{sec:model_description}, we model the formation of the confined CSM, and describe the temporal evolution of the shock, focusing on the formation of the CDS inside the shocked region. In Section \ref{sec:results}, we show the results of dust formation including the comparison of our model with the observation of SN IIn~1998S, which is suggested to have a confined CSM. In Section \ref{sec:conclusion}, we discuss the implications of our findings, including the survivability of dust grains and the possible contamination of kilonova (KN) surveys by IR dust emission. We conclude by outlining several caveats and directions for future work.

\section{Model description}
\label{sec:model_description}
\begin{figure}[t]
\centering
\includegraphics[width=\linewidth]{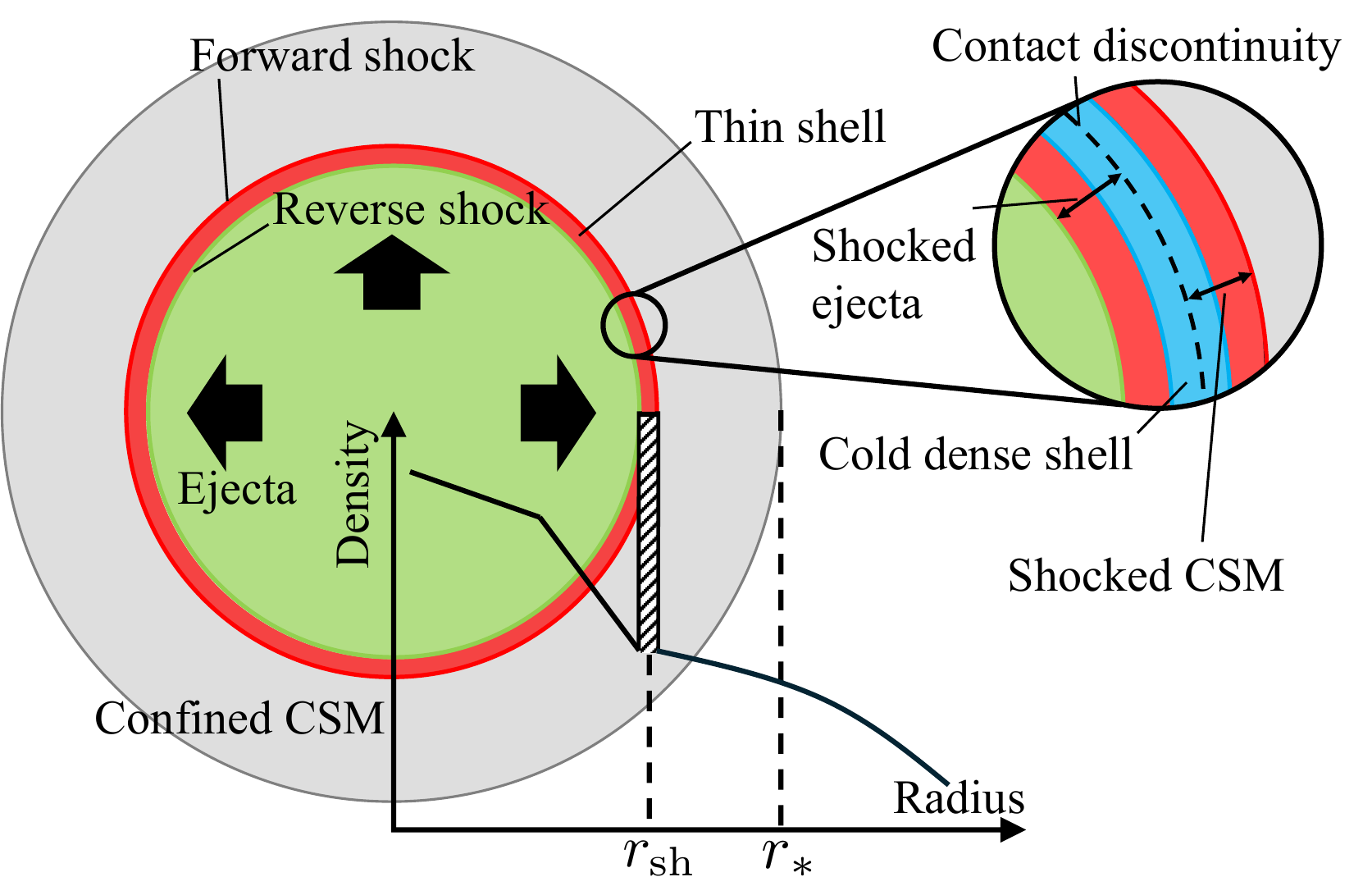}
    \caption{Schematic illustration of the interaction between the homologously expanding SN ejecta and the CSM (not scaled). This leads to the instantaneous formation of the CDS (blue-shaded region within the magnified view), where the new dust is expected to form. }
\label{fig:ponchie}
\end{figure}
We model newly formed dust in SNe II exploding inside confined CSM through the following procedure: First, we compute the mass eruption from a red supergiant (RSG) to construct the confined CSM.
Next, we calculate the evolution of the shock generated by the SN ejecta-CSM interaction using a thin-shell approximation. Finally, we estimate the mass of the CDS, where dust formation is expected to occur.
Throughout this study, we assume spherical symmetry of the system.
We present a schematic illustration in Figure \ref{fig:ponchie} that depicts the interaction between the SN ejecta and the CSM, which leads to the formation of a geometrically thin shell.

\subsection{Progenitor and Ejecta Structure}
We adopt the same progenitor model as in \citet{Takei_et_al_2022}, which corresponds to an RSG with a zero-age main sequence (ZAMS) mass of $M_\mathrm{ZAMS}=16\,M_\odot$, and metallicity of $Z=Z_\odot=0.014$ \citep{Asplund_et_al_2009}.
This progenitor was evolved until CC using the one-dimensional stellar evolution code \texttt{MESA} version 12778 \citep{Paxton_11,Paxton_13,Paxton_15,Paxton_18,Paxton_19,Jermyn_23}. The resultant mass of the progenitor is reduced to $14.7\,M_\odot$ due to the steady stellar wind prescribed by the standard ``dutch" wind scheme \footnote{\modify{The inlists used to create the progenitor model are available on Zenodo under an open Creative Commons Attribution license: \dataset[10.5281/zenodo.17034624]{https://doi.org/10.5281/zenodo.17034624}.}}. This progenitor mass is further reduced after the mass ejection, as we describe in Section \ref{sec:construct_CSM}.
In this work, the explosion energy $E_\mathrm{ej}$ is fixed at $10^{51}\,{\rm erg}$ unless otherwise mentioned.

The density structure of the homologously expanding SN ejecta ($v_\mathrm{ej}=r/t$) is modeled using a broken power-law profile following the model of \citet{Matzner_McKee_1999}. The density $\rho_\mathrm{ej}(r,\,t)$ is expressed as \citep[see also][]{Moriya_et_al_13},
\begin{eqnarray}
\rho_{\rm ej} (r,\,t)
&=& \left\{\begin{array}{ll}
t^{-3}\left[r/(gt)\right]^{-n} & (r/t > v_t),\\
t^{-3}(v_t/g)^{-n} \left[r/(tv_t)\right]^{-\delta}  & (r/t < v_t),
\end{array}\right.
\label{eq:rho_ej}
\end{eqnarray}
where $g$ and $v_t$ are constants derived from mass and energy constraints,
\begin{eqnarray}
    & \displaystyle g^n=\frac{1}{4\pi(n-\delta)}\frac{[2(5-\delta)(n-5)E_\mathrm{ej}]^{(n-3)/2}}{[(3-\delta)(n-3)M_\mathrm{ej}]^{(n-5)/2}},\\
    & \displaystyle v_t=\left[\frac{2(5-\delta)(n-5)E_\mathrm{ej}}{(3-\delta)(n-3)M_\mathrm{ej}}\right]^{1/2}.
\end{eqnarray}
Here $M_\mathrm{ej},\,\delta\simeq0$--$1$, and $n$ denote the ejecta mass, and inner and outer exponents of the ejecta, respectively. The outer density slope $n$ depends on the progenitor, and is expected to be $\sim12$ for RSGs \citep{Matzner_McKee_1999}.
The inner slope $\delta$ is fixed to 1 in \texttt{CHIPS}.
$M_\mathrm{ej}$ is determined by Equation (4) of \citet{Takei_et_al_2022} after the simulation for the formation of the confined CSM.
We fit the pressure $(p)$ and density $(\rho)$ structure of the progenitor after the mass eruption with $p\propto\rho^{1+1/N_\mathrm{pol}}$ to obtain the value of $n$, where $N_\mathrm{pol}$ denotes the polytropic index, since the outer density slope of the progenitor is almost the same as $n$.
Then $n$ is determined as,
\begin{eqnarray}
    n=\frac{N_\mathrm{pol}+1+3\beta N_\mathrm{pol}}{\beta N_\mathrm{pol}},
\end{eqnarray}
where $\beta\sim0.19$ \citep[see][]{Matzner_McKee_1999}. These processes can be conducted by the module implemented in \texttt{CHIPS}.

\subsection{Constructing the confined CSM}
\label{sec:construct_CSM}
Here we assume that the confined CSM originates from an eruptive mass-loss driven by energy injection at the base of the stellar envelope shortly before CC. 
We simulate this process using the mass eruption part of \texttt{CHIPS}, which solves the one-dimensional Lagrangian radiation hydrodynamics equations \citep[for the mass eruption part, see also][]{Kuriyama_Shigeyama_2020}.

In the simulation, a thermal energy $f_\mathrm{inj}$, scaled with the envelope's binding energy, is injected over 1000\,s, much shorter than the dynamical timescale of the envelope, and the temporal evolution of the CSM is followed for a duration $t_\mathrm{inj}$ before the explosion. However, for cases with $f_\mathrm{inj}<1$, fallback-induced shocks can introduce artificial structures near the interface between the CSM and progenitor \citep[see also Figure 2 of][]{Takei_et_al_2022}. To avoid this issue, we instead adopt an analytical form for the CSM density profile from \citet{Tsuna_et_al_21},
\begin{eqnarray}
    \rho_\mathrm{CSM}(r)=\rho_{*}\left[\frac{(r/r_{*})^{1.5/y_{*}}+(r/r_{*})^{n_\mathrm{out}/y_{*}}}{2}\right]^{-y_{*}},
    \label{eqn:CSM_prof}
\end{eqnarray}
where $\rho_*,\,y_*$, and $n_\mathrm{out}$ denote the density and the curvature at the transition point $r_*$, and the outer exponent of the CSM, respectively.
$n_\mathrm{out}$ is fixed to $12$, which is a typical power-law index for the outer envelope of RSGs \citep{Matzner_McKee_1999}, because the outer density profile of the CSM is shown to reflect the density gradient of the progenitor's outer layer like the ejecta from SNe \citep[e.g.,][]{Kuriyama_Shigeyama_2020,Tsuna_Takei_Shigeyama_2023,Tsuna_Takei_2023}. We determine the other three parameters $\rho_*,\,r_*$, and $y_*$ by fitting the above equation to the \texttt{CHIPS} output.

\subsection{Shock Evolution}
After the SN explosion, the homologously expanding ejecta collides with the confined CSM, generating a forward shock propagating into the CSM and a reverse shock moving back into the ejecta. These shocks heat the SN ejecta and CSM, and produce a thin shocked region with a radiative cooling timescale $\tau_\mathrm{cool}$.

Assuming that the shell width is negligible compared to the shock radius $r_\mathrm{sh}$, we adopt the thin-shell approximation \citep{Moriya_et_al_13}. The mass and momentum conservations of the shell are, respectively, described as,
\modify{
\begin{align}
    \frac{dM_\mathrm{sh}}{dt} &= 4\pi r_\mathrm{sh}^{2}
    \left[\rho_\mathrm{ej}(u_\mathrm{sh}-v_\mathrm{ej})
    +\rho_\mathrm{CSM}(u_\mathrm{sh}-v_\mathrm{CSM})\right], 
    \label{eqn:mass_consv} \\
    M_\mathrm{sh}\frac{du_\mathrm{sh}}{dt} &= 4\pi r_\mathrm{sh}^{2}
    \left[\rho_\mathrm{ej}(u_\mathrm{sh}-v_\mathrm{ej})^{2}
    -\rho_\mathrm{CSM}(u_\mathrm{sh}-v_\mathrm{CSM})^{2}\right],
    \label{eqn:mom_consv}
\end{align}}
where $u_\mathrm{sh}=dr_\mathrm{sh}/dt$ denotes the velocity of the shell, $v_\mathrm{CSM}$ is the velocity of the CSM, and $M_\mathrm{sh}$ is the mass of the shocked region.
These equations are integrated numerically from $t=0.2\,{\rm d}$ with a fourth-order Runge-Kutta method. Since the inner part of the CSM described by Equation (\ref{eqn:CSM_prof}) follows $\rho_\mathrm{CSM}\propto r^{-1.5}$, the initial conditions on $r_\mathrm{sh},\,u_\mathrm{sh}$, and $M_\mathrm{sh}$ are determined from the analytical solution derived in \citet{Moriya_et_al_13}.

\subsection{Formation and Evolution of the cold dense shell}
The shock-heated gas behind both the reverse and forward shocks can cool efficiently via radiation, leading to the formation of a CDS near the contact discontinuity as illustrated in Figure \ref{fig:ponchie}.
The evolution of the CDS mass $M_\mathrm{CDS}(t)$ can be crudely estimated by,
\begin{eqnarray}
    & \displaystyle \frac{dM_\mathrm{CDS}}{dt}\sim \frac{M_\mathrm{hot}}{\tau_\mathrm{cool}},\\
    & M_\mathrm{CDS}=M_\mathrm{sh}-M_\mathrm{hot},
\end{eqnarray}
where $M_\mathrm{hot}$ denotes the still-hot gas in the shocked region.
We integrate the above equation until the shock reaches $r_\mathrm{sh}=r_*$.
Under the assumption of adiabatic shock heating, the post-shock gas temperature can be determined from the Rankine-Hugoniot relations,
\begin{eqnarray}
    T_\mathrm{gas}&=&\frac{3}{16}\frac{\mu m_\mathrm{u}}{k}v_\mathrm{d}^{2},\nonumber \\
    &\approx &1.4\times10^7\,{\rm K}\left(\frac{v_\mathrm{d}}{10^{8}\,{\rm cm\>s^{-1}}}\right)^{2},
\end{eqnarray}
where $\mu,\,k$, and $m_\mathrm{u}$ denote a mean molecular weight, Boltzmann constant, and unified atomic mass unit, respectively.
$v_\mathrm{d}$ is the velocity of the post-shock gas in the rest frame of each shock, i.e., $v_\mathrm{d}=r_\mathrm{sh}/t-u_\mathrm{sh}$ for the reverse shock, and $v_\mathrm{d}=u_\mathrm{sh}-v_\mathrm{CSM}$ for the forward shock. At temperatures exceeding $\sim10^{7}$--$10^{8}\,{\rm K}$, radiative cooling is predominantly governed by free-free emission. The corresponding cooling timescale can be thus approximated as,
\begin{eqnarray}
    \tau_\mathrm{cool}\sim \frac{U_\mathrm{int}}{4\pi\eta_\mathrm{ff}},
\end{eqnarray}
where $U_\mathrm{int}$ and $\eta_\mathrm{ff}\,[{\rm erg\>s^{-1}cm^{-3}\>sr^{-1}}]$ denote the internal energy of the post-shock gas and the frequency-integrated free-free emissivity, respectively.

After the shock passes the transition point in Equation (\ref{eqn:CSM_prof}), $r_\mathrm{sh}\gtrsim r_*$, the shocked region expands nearly freely, and the gas inside the shock cools adiabatically. As the temperature of the gas $T_\mathrm{gas}$ decreases further, it can eventually reach conditions favorable for dust condensation. In the adiabatic regime, $T_\mathrm{gas}$ approximately follows, 
\begin{eqnarray}
    T_\mathrm{gas} \propto r_\mathrm{sh}^{-1},
\end{eqnarray}
for a radiation-dominated shell, while the gas density of the CDS $\rho_\mathrm{CDS}$ decreases as,
\begin{eqnarray}
    \rho_\mathrm{CDS} \propto r_\mathrm{sh}^{-3}.
\end{eqnarray}
This cooling process differs from that in \citet{Sarangi_Slavin_2022}, where the gas is assumed to cool down to the condensation temperature through radiative processes described by the cooling function extracted from the spectral synthesis code \texttt{CLOUDY} \citep{Ferland_et_al_2013}. 
While the more extended wind that has a power-law distribution $(\rho\propto r^{-2})$ is considered in their work, the sharp drop in CSM density near the outer edge leads to the termination of the interaction between the ejecta and the CSM in our model. This transition initiates the free expansion of the CDS, during which the gas undergoes rapid adiabatic cooling and reaches the condensation temperature on a much shorter timescale.

At $r_\mathrm{sh}\lesssim r_*$, the Rankine-Hugoniot relations provide an estimate of the pressure within the CDS $p_\mathrm{CDS}$ under the assumption that the CDS is in pressure balance. If we further assume that the CDS has cooled to $T_\mathrm{CDS}\sim10^{4}\,{\rm K}$ according to the cooling function \citep{Sutherland_Dopita_1993}, $\rho_\mathrm{CDS}$ can be determined from the equation of state below,
\begin{eqnarray}
    p_\mathrm{CDS}=\frac{1}{3}a_r T_\mathrm{CDS}^{4}+\frac{\rho_\mathrm{CDS}}{\mu m_\mathrm{u}}kT_\mathrm{CDS}, \label{eqn:eos}
\end{eqnarray}
where $a_r$ denotes the radiation constant. We adopt a mean molecular weight of $\mu\approx0.62$, which corresponds to a fully ionized gas with solar abundance.
Equation (\ref{eqn:eos}) is used to evaluate $\rho_\mathrm{CDS}$ at the onset of the free expansion.

\subsection{Estimating Dust Luminosity}
\label{sec:estimating_dust_luminosity}
We calculate the dust luminosity assuming that all metals in the CDS condense into dust, and that dust grains form instantaneously once the CDS temperature falls below the condensation threshold,
\begin{eqnarray}
T_\mathrm{d,\,max}=2000\,{\rm K},
\end{eqnarray}
which is a typical threshold for carbon to condense into amorphous
carbon grains \citep{Todini_Ferrara_2001}.
If the dust mass $M_\mathrm{d}=M_\mathrm{d}(Z)$ (which depends on the metallicity) and temperature $T_\mathrm{d}$ are given, the monochromatic luminosity at wavelength $\lambda$ can be expressed as,
\begin{equation}
    L_\mathrm{d,\,\lambda} = 4\pi M_\mathrm{d}^\mathrm{obs}\, \bar{\kappa}_\lambda\, B_\lambda(T_\mathrm{d}),\label{eqn:dust_lum_mono}
\end{equation}
where $M_\mathrm{d}^\mathrm{obs}$ is the observable dust mass, $\bar{\kappa}_\lambda$ is the dust mass absorption coefficient, and $B_\lambda(T_\mathrm{d})$ is the Planck function at dust temperature $T_\mathrm{d}$.
It should be noted that $M_\mathrm{d}^\mathrm{obs}(\lambda,\,t)$ depends on $\lambda$ and is generally smaller than the actual dust mass $M_\mathrm{d}$ due to the finite optical depth of the dust shell $\tau_\lambda$. According to \citet{Dwek_et_al_2019}, the observable mass is related to the total dust mass through the photon escape probability $P_\mathrm{esc}(\tau_\lambda)$ as,
\begin{eqnarray}
    M_\mathrm{d}^\mathrm{obs} = P_\mathrm{esc}(\tau_\lambda)\, M_\mathrm{d},
\end{eqnarray}
where 
$P_\mathrm{esc}(\tau_\lambda)$ for IR photons from a dusty medium with optical depth $\tau_\lambda$ is given by,
\begin{eqnarray}
    & \displaystyle P_\mathrm{esc}(\tau_\lambda) = \frac{3}{4\tau_\lambda} \left[1 - \frac{1}{2\tau_\lambda^{2}} + \left(\frac{1}{\tau_\lambda} + \frac{1}{2\tau_\lambda^{2}}\right)e^{-2\tau_\lambda} \right], \\
    & \displaystyle \tau_\lambda=\frac{\bar{\kappa}_\lambda M_\mathrm{d}}{4\pi r_\mathrm{sh}^{2}},
\end{eqnarray}
as derived under the assumption of a homogeneous, spherically symmetric dust distribution \citep{Cox_Mathews_1969,Osterbrock_Ferland_2006}.
Although our model does not assume a uniformly filled sphere but instead considers dust confined within a geometrically thin shell, we adopt the same expression for $P_\mathrm{esc}(\tau_\lambda)$ as an approximation. This treatment has been used in previous studies of dust emission in thin-shell geometries \citep{Shahbandeh_et_al_2025,Tinyanont_et_al_2025} and provides a reasonable estimate of the escaping fraction, especially when the shell is optically thick or moderately thin.

The mass absorption coefficient $\kappa_\lambda$ is a function of the grain size $a$, the mass density of a dust particle $\rho_\mathrm{gr}$, and the emission/absorption efficiency $Q_\lambda(a)$ at wavelength $\lambda$ described as \citep{Dwek_1983},
\begin{eqnarray}
    & \displaystyle \kappa_\lambda(a)=\frac{3Q_\lambda(a)}{4a \rho_\mathrm{gr}},\\
    & \displaystyle \bar{\kappa}_\lambda=\frac{\int_{a_\mathrm{min}}^{a_\mathrm{max}} \kappa_\lambda(a)m_\mathrm{d}(a)n(a)da}{\int_{a_\mathrm{min}}^{a_\mathrm{max}} m_\mathrm{d}(a)n(a)da},\label{eqn:opacity}
\end{eqnarray}
where $n(a)$ and $m_\mathrm{d}(a)=(4\pi/3)a^{3}\rho_\mathrm{gr}$ denote the dust size distribution and the mass of a dust grain with size $a$, respectively. We set the distribution limits at $a_\mathrm{min}=0.005\,\mu{\rm m}$ and $a_\mathrm{max}=0.05\,\mu{\rm m}$, which is based on the detailed modeling of the dust formation within the SN ejecta of the He core \citep{Todini_Ferrara_2001,Nozawa_et_al_2003}.
To calculate the mass-averaged absorption coefficient $\bar{\kappa}_\lambda$ which is used to evaluate the dust luminosity from Equation (\ref{eqn:dust_lum_mono}), we need $\rho_\mathrm{gr},\,Q_\lambda(a)$ and $n(a)$.
For the dust mass density, we adopt $\rho_\mathrm{gr}=2.3\,{\rm g\>cm^{-3}}$, which is commonly used for graphite grain in the literature \citep[see e.g.,][]{Draine_Lee_1984,Laor_Draine_1993}\footnote{If we choose silicate for dust species, $\rho_\mathrm{gr}\sim3.3\,{\rm g\>cm^{-3}}$, which does not change the opacity much.}.
For $Q_\lambda(a)$, we use the data calculated based on \citet{Laor_Draine_1993}\footnote{We downloaded the data from the following website (Gra\_81.gz): \url{https://www.astro.princeton.edu/\string~draine/dust/dust.diel.html}}.
$n(a)$ is assumed to follow a power-law size distribution model,
\begin{eqnarray}
n(a)\propto a^{-3.5},\label{eqn:MRN}
\end{eqnarray}
which has been widely used to describe interstellar dust \citep*[][hereafter MRN]{Mathis_et_al_1977}. Although the actual size distribution of dust grains formed in SNe and their environments may differ, especially under high-density and non-equilibrium conditions, deriving a realistic distribution remains a major challenge. It requires detailed modeling of dust nucleation and growth in a highly time-dependent, radiation- and chemistry-coupled environment \citep[e.g.,][]{Kozasa_Hasegawa_1987,Nozawa_et_al_2003,Nozawa_Kozasa_2013}. Therefore, we adopt the MRN distribution as a reasonable approximation that allows for simplified yet consistent estimates of dust emission and opacity.

\section{Results}
\label{sec:results}
In this section, we present the results of our simulations of the interaction between SN ejecta and a confined CSM. We focus on the evolution of the shock and the formation of the CDS, followed by the estimation of the resulting dust mass and the IR emission from the newly formed dust. Finally, we compare the model predictions with observational data of SN~1998S.
The simulation are conducted by varying two key parameters (see Section \ref{sec:construct_CSM}): the injected energy scaled with the envelope's binding energy $f_\mathrm{inj}$ and the time from CSM ejection to CC $t_\mathrm{inj}$. These parameters affect the mass and spatial extent of the CSM; larger $f_\mathrm{inj}$ results in more massive CSM, while longer $t_\mathrm{inj}$ leads to more extended and dilute CSM.
We explore a range of $f_\mathrm{inj}=0.3,\,0.4,\,0.5,\,0.6,\,0.8$ and $t_\mathrm{inj}=3,\,4,\,\cdots,15\,{\rm yr}$.
For each parameter set, we track the shock dynamics, the formation and evolution of the CDS, and the resulting dust formation and emission.

\subsection{Evolution of the Shock and Shocked Mass}
\begin{figure*}
\centering
\includegraphics[width=\linewidth]{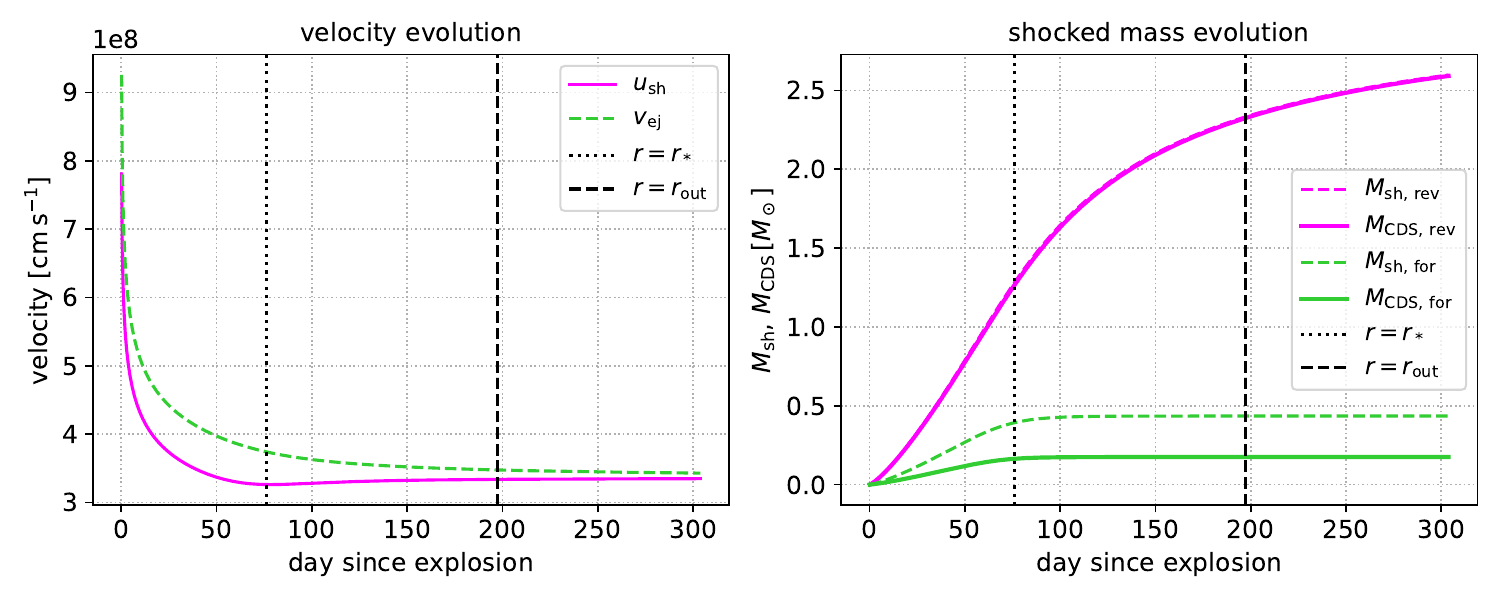}
    \caption{The temporal evolution of the shock velocity, ejecta velocity, shocked mass, and CDS mass. 
    \modify{The time at which the shock reaches the transition radius and the outer edge of the CSM is marked by the vertical lines $r_*$ and $r_\mathrm{out}$, respectively.}
    The subscript rev (for) denotes the mass of the shocked region/CDS on the shocked ejecta (CSM) side. The adopted parameters are $t_\mathrm{inj}=10\,{\rm yr}$ and $f_\mathrm{inj}=0.5$.}
\label{fig:shock_evolution}
\end{figure*}
We plot in Figure \ref{fig:shock_evolution} the temporal evolution of the velocities of the shock $u_\mathrm{sh}$ and the unshocked ejecta $v_\mathrm{ej}$ at the shock position (left panel), as well as the masses of the shocked shell $M_\mathrm{sh,\,rev},\,M_\mathrm{sh,\,for}$ and CDS $M_\mathrm{CDS,\,rev},\,M_\mathrm{CDS,\,for}$, where the subscripts rev/for denote the mass of the shocked region and CDS on the shocked ejecta/CSM side, respectively (right panel). 
Until the shock reaches $r_\mathrm{sh}=r_*$, the shock velocity $u_\mathrm{sh}$ decreases due to the relatively shallow density gradient of the inner CSM, characterized by a slope of $-1.5$, which is flatter than $-3$.
Once the shock propagates beyond $r_\mathrm{sh}=r_*$, the steep decline in the CSM density leads to an acceleration of the shock front, which is also mentioned in \citet{Chevalier_1982}. 
Consequently, the velocity of the unshocked ejecta at the shock position $v_\mathrm{ej}=r_\mathrm{sh}/t$ approaches that of the shock itself $u_\mathrm{sh}$, resulting in a gradual deceleration of shell growth, as shown in the right panel.

The right panel of Figure \ref{fig:shock_evolution} also shows that the evolution of the CDS mass differs significantly between the ejecta side (reverse shock) and the CSM side (forward shock).
This difference arises from the large contrast between the densities of the ejecta and the CSM, which leads to a significantly shorter cooling timescale on the ejecta side. The larger mass of the CDS on the ejecta side significantly contributes to the total dust formation, making this region the primary site of dust production.

In the parameter space explored in this work, the reverse shock does not reach the ejecta of the He core, and therefore the shocked region primarily consists of the hydrogen-rich envelope, which has a relatively low metallicity. However, if a much larger energy of $f_\mathrm{inj}\gg1$ is injected, the entire envelope can be expelled from the progenitor. In such a case, the reverse shock would propagate deeper into the ejecta and eventually reach the He core or even the inner CO core. This would significantly increase the metallicity in the shocked region, allowing a larger fraction of metals to be available for dust formation. Such scenarios are beyond the scope of this work, but may be relevant for understanding the upper range of dust yields in some extreme events.

\subsection{Expected Dust Mass}
\label{sec:expected_dust_mass}
\begin{figure*}
\centering
\includegraphics[width=\linewidth]{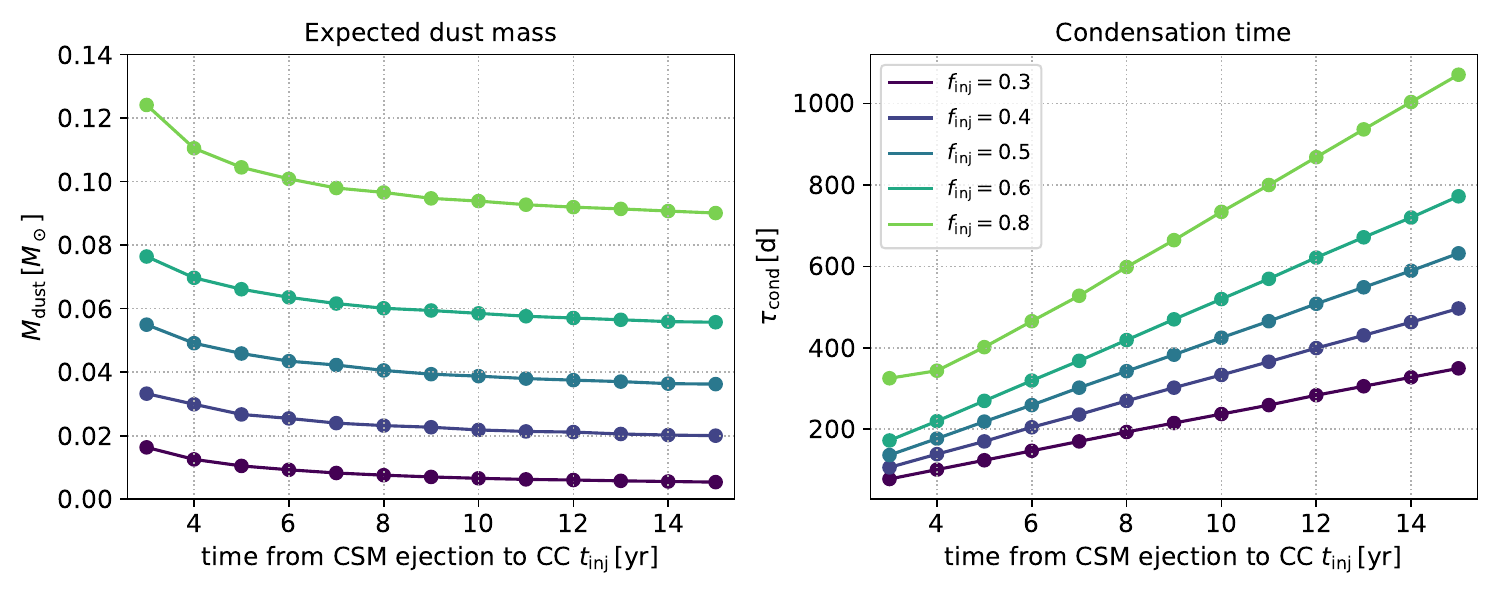}
    \caption{The expected dust mass formed in the CDS (left panel) and the condensation time of dust $\tau_\mathrm{cond}$ (right panel) as a function of $t_\mathrm{inj}$ for each value of $f_\mathrm{inj}$.}
\label{fig:dust_mass}
\end{figure*}
Under the assumption that all metals contained in the CDS condense into dust, we calculate the total dust mass and plot it as a function of $t_\mathrm{inj}$ in the left panel of Figure \ref{fig:dust_mass} for each value of $f_\mathrm{inj}$.

The increase in the resulting dust mass with increasing $f_\mathrm{inj}$ is primarily due to the enhanced mass of the CSM, which slows down the forward shock propagation. Because of this deceleration, the shock sweeps up a larger amount of the outer ejecta before reaching the radius $r=r_*$. Consequently, the mass of the CDS becomes larger, providing more material for dust formation.
In contrast, increasing $t_\mathrm{inj}$ results in a smaller dust mass. This is because a longer $t_\mathrm{inj}$ leads to a more dilute CSM, allowing the shock to propagate more rapidly. As a consequence of the higher shock velocity, less ejecta is swept up before the shock reaches $r=r_*$, leading to a smaller mass accumulated in the CDS.

In the right panel of Figure \ref{fig:dust_mass}, we show the condensation time of dust $\tau_\mathrm{cond}$ when the gas temperature reaches $T_\mathrm{d,\,max}$.
As can be seen from this figure, $\tau_\mathrm{cond}$ increases monotonically with both $t_\mathrm{inj}$ and $f_\mathrm{inj}$. A longer $t_\mathrm{inj}$ corresponds to a more extended CSM, which increases the time required for the shock to sweep up the CSM and, consequently, leads to a delayed onset of free expansion. In addition, larger $f_\mathrm{inj}$ implies a greater energy budget available to expel the CSM, leading to a higher CSM velocity at $r=r_*$ for a fixed value of $t_\mathrm{inj}$, and thus a more extended CSM. As noted above, the higher density of the CSM in cases with larger $f_\mathrm{inj}$ further reduces the shock velocity. These effects combine to produce a monotonic increase in $\tau_\mathrm{cond}$ with increasing $f_\mathrm{inj}$.

\begin{figure*}
\centering
\includegraphics[width=\linewidth]{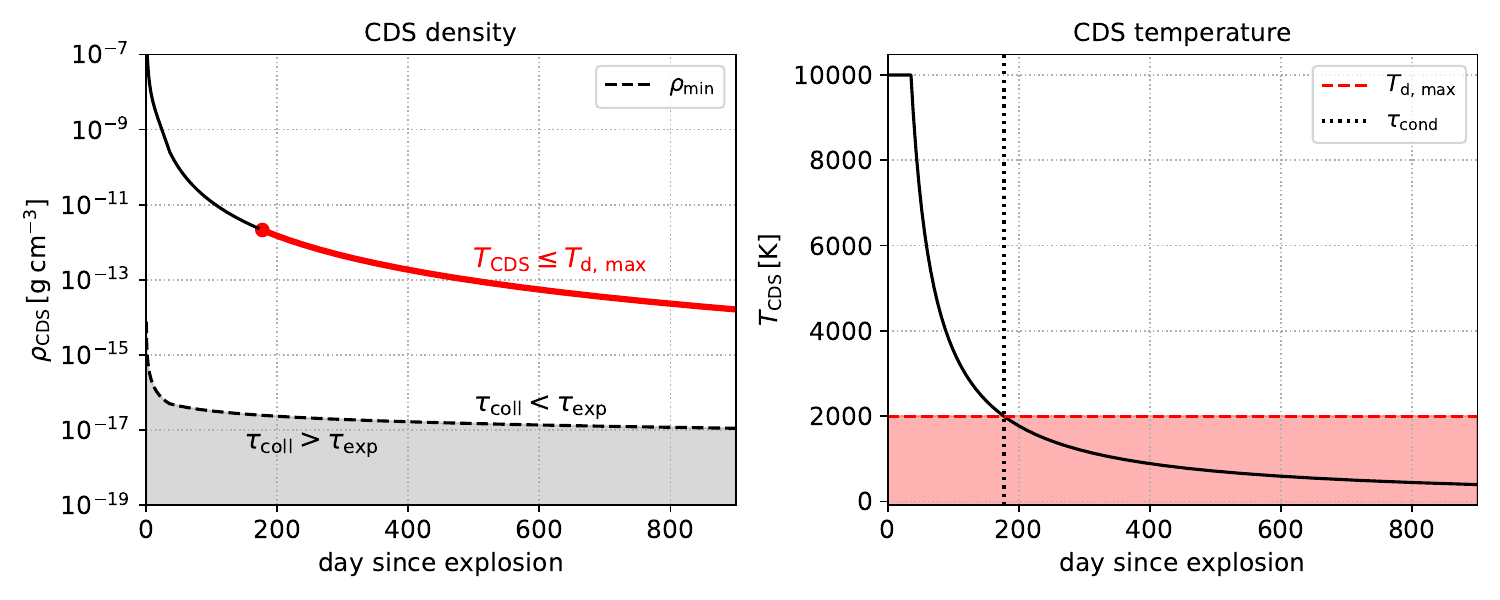}
    \caption{The temporal evolution of the density (left panel) and temperature (right panel) of CDS. 
    The gray-shaded area in the left panel represents the region in which $\tau_\mathrm{coll}>\tau_\mathrm{exp}$, suggesting that dust formation is suppressed under these conditions. The red-shaded region in the right panel indicates the area where $T\leq T_\mathrm{d,\,max}$, representing conditions under which dust formation is possible. After sweeping through the confined CSM, the CDS undergoes rapid adiabatic cooling, causing its temperature to drop sharply from $\sim10^{4}\,{\rm K}$, which can be seen from the right panel.
    The adopted parameters are $f_\mathrm{inj}=0.3$ and $t_\mathrm{inj}=10\,{\rm yr}$ (see Figure \ref{fig:dust_mass}).}
\label{fig:CDS_property}
\end{figure*}

In estimating the expected dust mass, it is important to assess whether the physical conditions allow for dust formation. A necessary condition is that the timescale of the collision between atoms $\tau_\mathrm{coll}\sim(n_i\sigma\braket{v_\mathrm{th}})^{-1}$ is much shorter than the expansion timescale $\tau_\mathrm{exp}\sim r_\mathrm{sh}/u_\mathrm{sh}$, i.e., 
\begin{eqnarray}
\tau_\mathrm{coll}\ll \tau_\mathrm{exp},
\end{eqnarray}
where $n_i$, $\sigma$, and $\braket{v_\mathrm{th}}$ denote the number density and collision cross section of the element $i$, and mean thermal velocity, respectively. Substituting $n_i=X_i\rho_\mathrm{CDS}/A_i m_\mathrm{u},\,\sigma=\pi a_i^2=\pi a_0^2 A_i^{2/3},\,\braket{v_\mathrm{th}}=(2kT_\mathrm{CDS}/A_i m_\mathrm{u})^{1/2}$ into $\tau_\mathrm{coll}$ yields \citep[see also][]{Takami_et_al_2014},
\begin{eqnarray}
    \tau_\mathrm{coll}&\sim&\frac{(A_im_\mathrm{u})^{3/2}}{\pi a_0^2 A_i^{2/3}X_i\rho_\mathrm{CDS}(2kT_\mathrm{CDS})^{1/2}} \nonumber \\
    &\sim&80\,{\rm s}\left(\frac{T_\mathrm{CDS}}{2000\,{\rm K}}\right)^{-1/2}\left(\frac{\rho_\mathrm{CDS}}{10^{-12}\,{\rm g\>cm^{-3}}}\right)^{-1} \nonumber \\
    &\times&\left(\frac{X_i}{10^{-3}}\right)^{-1}\left(\frac{A_i}{12}\right)^{5/6}\left(\frac{a_0}{1\,\mathrm{\AA}}\right)^{-2},
\end{eqnarray}
where $a_i,\,A_i,\,X_i$, and $a_0$ denote the radius, molecular weight, mass fraction of the element $i$, and normalization radius, respectively.
Assuming $X_\mathrm{C}=10^{-3}$, which is a typical value for the abundances in the hydrogen-rich envelope of RSGs \citep{Davies_Dessart_2019}, this timescale is found to be significantly shorter than $\tau_\mathrm{exp}\sim r_\mathrm{sh}/u_\mathrm{sh}\sim t$ after the condensation time $\tau_\mathrm{cond}$ (Figure \ref{fig:dust_mass}). Given that the ratio $\tau_\mathrm{cond}/\tau_\mathrm{coll}$ is $\sim10^{5}$, we estimate that an atom can experience $\sim10^{5}$ collisions before the gas becomes too diffuse, allowing sufficient time for aggregation into dust grains.

To further elucidate the physical conditions governing dust formation we present in Figure \ref{fig:CDS_property} the temporal evolution of $\rho_\mathrm{CDS}$ and $T_\mathrm{CDS}$ for a representative model, $f_\mathrm{inj}=0.3$ and $t_\mathrm{inj}=10\,{\rm yr}$. These quantities directly affect $\tau_\mathrm{coll}$, and thus play a key role in determining whether the condition $\tau_\mathrm{coll}<\tau_\mathrm{exp}$ is satisfied during the evolution.
We select a model with relatively low CSM density, which provides a conservative estimate of the dust formation conditions. If the criterion is satisfied in such environments, it is expected to hold more robustly in models with denser CSM.
As shown in the figure, the gas rapidly cools and remains at sufficiently high density, such that $\tau_\mathrm{coll}<\tau_\mathrm{exp}$ is satisfied throughout the evolution, indicating favorable conditions for dust condensation.

\subsection{Infrared Emission from the Newly-Formed Dust}
\label{sec:IR_emission_from_dust}
\begin{figure}[t]
\centering
\includegraphics[width=\linewidth]{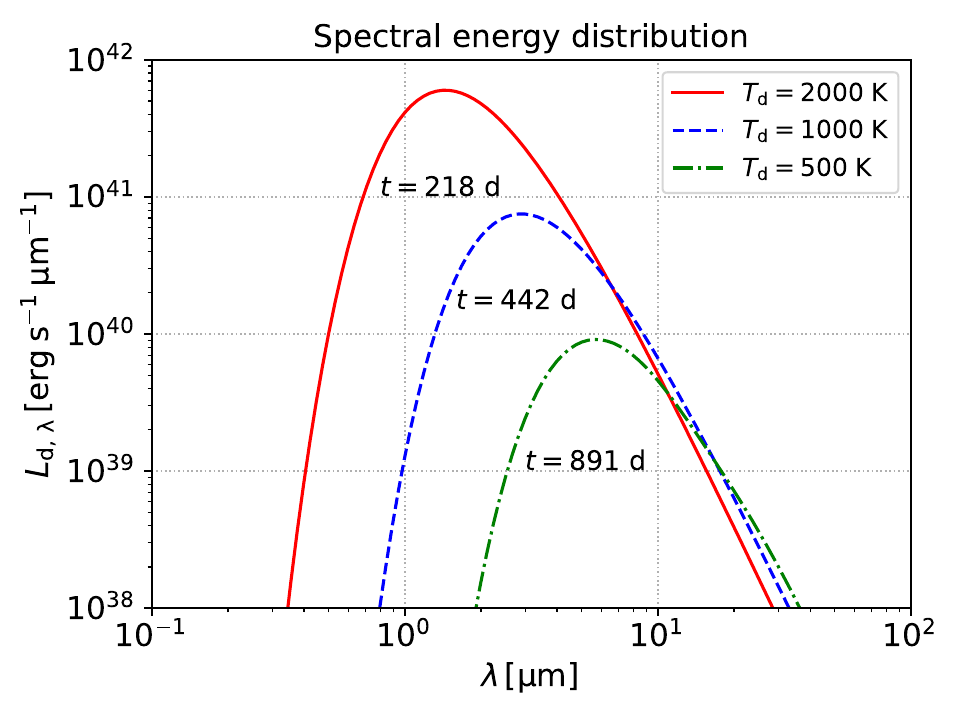}
    \caption{SEDs of the thermal emission from newly-formed dust at several epochs $t=218,\,442,\,891\,{\rm d}$, which correspond to $T_\mathrm{d}=2000,\,1000,\,500\,{\rm K}$, respectively. The adopted parameters are $f_\mathrm{inj}=0.5$ and $t_\mathrm{inj}=5\,{\rm yr}$.}
\label{fig:SED}
\end{figure}
In this section, we present the IR emission from the newly-formed dust. We focus on the spectral energy distribution and the total IR luminosity, both of which serve as key diagnostics of the dust mass and temperature.
Here the dust temperature $T_\mathrm{d}$ is assumed to follow the gas temperature $T_\mathrm{gas}$ in the CDS.

Figure \ref{fig:SED} displays the spectral energy distributions (SEDs) of the thermal emission from the dust at several representative epochs. The emission peaks in the near- to mid-IR range, reflecting the cooling of the dust as the CDS expands. Initially, the peak is located around $\lambda\sim1.4\,{\rm \mu m}$, following Wien's displacement law. As the temperature decreases, this peak shifts to longer wavelength.

\begin{figure}[t]
\centering
\includegraphics[width=\linewidth]{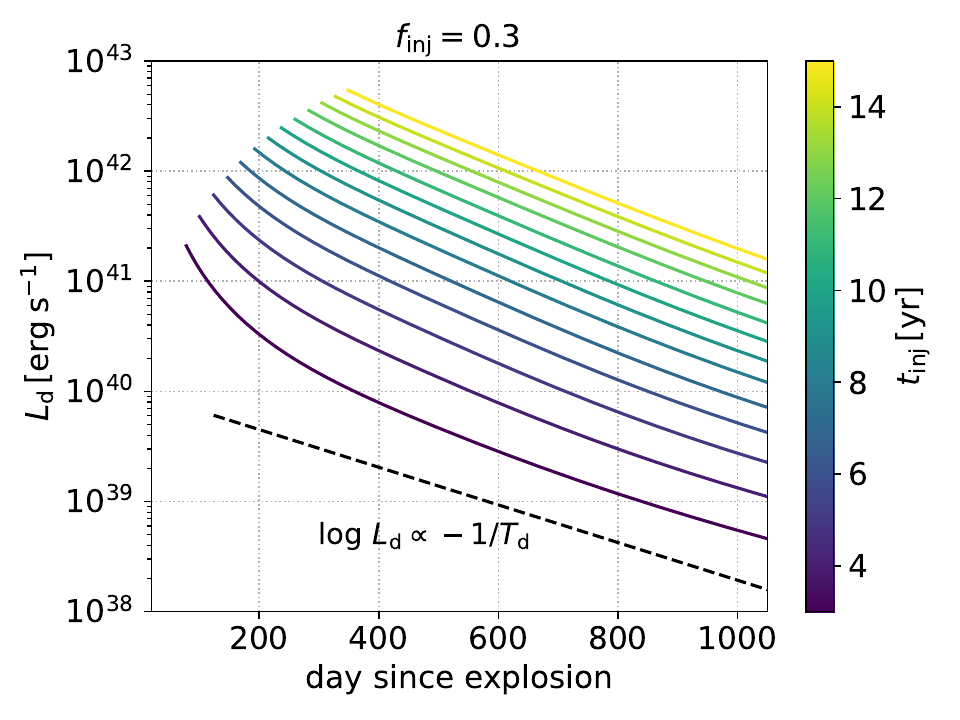}
    \caption{The expected light curves of dust assuming $T_\mathrm{d}=T_\mathrm{gas}$ for $t_\mathrm{inj}=3,\,4,\,\cdots,\,15\,{\rm yr}$. The dashed line indicates the dust luminosity which is proportional to $\log L_\mathrm{d}\propto -1/T_\mathrm{d}$. The adopted value for $f_\mathrm{inj}$ is 0.3.}
\label{fig:dust_luminosity}
\end{figure}
We calculate the total dust luminosity $L_\mathrm{d}$ by integrating Equation (\ref{eqn:dust_lum_mono}) from $\lambda_\mathrm{min}=10^{-3}\,{\rm \mu m}$ to $\lambda_\mathrm{max}=10^{3}\,{\rm \mu m}$, 
\begin{eqnarray}
L_\mathrm{d}=\int_{\lambda_\mathrm{min}}^{\lambda_\mathrm{max}}L_\mathrm{d,\,\lambda}d\lambda,
\end{eqnarray}
where $\lambda_\mathrm{min}$ and $\lambda_\mathrm{max}$ are the lower and upper bounds of the data points given in the table of the emissivity $Q_\lambda(a)$ that we use to calculate the opacity. We plot $L_\mathrm{d}$ in Figure \ref{fig:dust_luminosity} for each $t_\mathrm{inj}$.
From this figure, we can see the monotonic increase of the luminosity with increasing $t_\mathrm{inj}$. 
This is attributed to the increase in $M_\mathrm{d}^\mathrm{obs}$, which results from the decreasing optical depth at larger shock radii.
In particular, models with longer $t_\mathrm{inj}$ exhibit a delayed onset of the light curve, as the transition to the free-expansion phase occurs at later times due to the prolonged interaction between the ejecta and the extended CSM.

We also find that, at later times, $L_\mathrm{d}$ is roughly proportional to,
\begin{eqnarray}
    \log L_\mathrm{d}\propto -\frac{1}{T_\mathrm{d}}.
\end{eqnarray}
This behavior primarily arises because the emissivity is much lower at longer wavelengths, and only the Wien tail on the short wavelength side of the Planck function contributes significantly. This relation can be used to estimate the onset of free expansion, that is, the radial extent of the CSM, based on the temporal evolution of the temperature.

Although the dust forms at a specific epoch, the observed IR light curve does not exhibit an instantaneous rise. This is due to the finite light-travel time across the emitting shell, as photons from the far side of the shell arrive later than those from the near side. As a result, the rising phase of the light curve is smeared over a timescale of $\sim r_\mathrm{sh}/c$, which is typically $\sim1$--$10\,{\rm d}$. Similar effects have been discussed in previous studies of IR emission from SNe \citep[e.g.,][]{Mattila_et_al_2008,Fox_et_al_2011}, where light-travel delays lead to a temporal smearing of the light curve.

\subsection{Comparison with Observation}
\begin{figure*}[t]
\centering
\includegraphics[width=\linewidth]{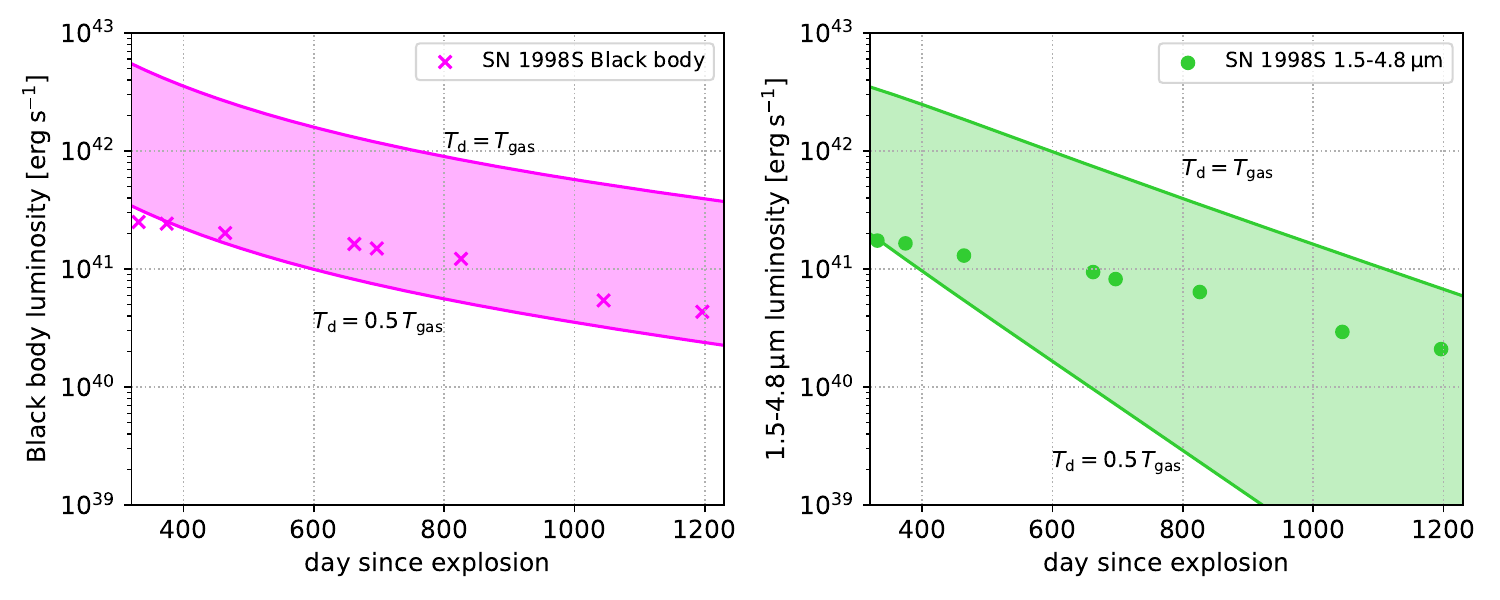}
    \caption{Comparison of the dust IR luminosity with our model. We assume $T_\mathrm{d}=T_\mathrm{gas}$ when we calculate the upper limit, while $T_\mathrm{d}=0.5\,T_\mathrm{gas}$ for lower limit. Left: Blackbody luminosity. Right: IR luminosity integrated over 1.5--4.8${\rm \,\mu m}$. The data were taken from \citet{Pozzo_et_al_2004}.}
\label{fig:cmp_SN1998S}
\end{figure*}
Here we compare our dust modeling with the observation of SN~1998S, which was discovered on 1998 March 2.68 UT in NGC 3877 \citep{Li_et_al_1998}.
This SN is an SN II with narrow hydrogen emission lines at early epoch, suggesting the presence of the confined CSM \citep[e.g.,][]{Fassia_et_al_2001}. The IR excess was observed at $\sim300$\,d after its explosion and the newly-formed dust mass is estimated to be at least $10^{-3}\,M_\odot$ \citep{Pozzo_et_al_2004}. 
\citet{Takei_et_al_2022} derived the model parameters, $E_\mathrm{ej}=2.5\times10^{51}\,{\rm erg},\,M_\mathrm{ZAMS}=20\,M_\odot,\,f_\mathrm{inj}=0.7,\,t_\mathrm{inj}=11\,{\rm yr}$, by fitting the optical light curve of SN~1998S with a model constructed using \texttt{CHIPS}. In this section, we investigate whether the same parameter set can also reproduce the light curve of dust. Given the uncertainties associated with the parameter estimation, it is reasonable to explore nearby regions of the parameter space to evaluate the robustness of the model.

Figure \ref{fig:cmp_SN1998S} shows a comparison between the light curve calculated with our dust emission model using Equation (\ref{eqn:dust_lum_mono}), (\ref{eqn:opacity}) and the IR emission attributed to dust formation in SN~1998S, varying $T_\mathrm{d}$ from $0.5\,T_\mathrm{gas}$ to $T_\mathrm{gas}$. 
For this comparison, we slightly adjusted the parameters from the best-fit parameters of \citet{Takei_et_al_2022}, changing $t_\mathrm{inj}$ from 11\,yr to 8\,yr and $E_\mathrm{ej}$ from $2.5\times10^{51}\,{\rm erg}$ to $3\times10^{51}\,{\rm erg}$.
As can be seen from the figure, the luminosity from dust observed in SN~1998S is broadly consistent with our model. The left panel displays the bolometric luminosity $L_\mathrm{d}$. Since the luminosity roughly scales as $T_\mathrm{d}^{4}$, the light curve assuming $T_\mathrm{d}=0.5\,T_\mathrm{gas}$ is reduced to $1/16$ of that for $T_\mathrm{d}=T_\mathrm{gas}$. In the right panel, we plot the luminosity integrated from $\lambda=1.5\,{\rm \mu m}$ to $\lambda=4.8\,{\rm \mu m}$. With the evolution of the system, the decreasing temperature of the CDS leads to a shift of the SED peak toward longer wavelengths. In the case of $T_\mathrm{d}=0.5\,T_\mathrm{gas}$, the observed wavelengths fall further into the Wien tail of the blackbody spectrum, resulting in a more significant reduction in luminosity.
In our model, the total mass of dust formed is estimated to be approximately $10^{-2}\,M_\odot$, which is broadly consistent with the lower limit inferred by \citet{Pozzo_et_al_2004} based on IR observations of SN~1998S. In their analysis, the dust mass was estimated by comparing the observed IR flux with optically thin blackbody emission from dust. This assumption implies that their estimate represents a lower limit, since optical depth effects can obscure a significant fraction of the emission.

\section{Discussion and Conclusion}
\label{sec:conclusion}
In this work, we have constructed a self-consistent model that follows the formation of confined CSM and its subsequent interaction with SN ejecta, enabling us to track the evolution of the resulting CDS.
Based on the thermodynamic conditions in the CDS, we assess the potential for dust formation under various physical conditions of the CSM.
When the SN ejecta collides with the CSM, a radiative shock forms and compresses the gas into a dense shell. After the shocked region breaks out of the CSM, it enters a phase of adiabatic expansion, during which the gas cools rapidly. This sequence of compression and expansion facilitates the attainment of thermodynamic conditions suitable for dust condensation.
Using the open-source code \texttt{CHIPS}, we modeled a range of confined CSM configurations and demonstrated that CDS formation occurs robustly across a broad parameter space, particularly with respect to CSM mass and radial extent.
The resulting dust mass spans a wide range from $\sim10^{-3}\,M_\odot$ to $0.1\,M_\odot$, reflecting the diversity in CSM density and its geometric distribution. \modify{We note that these results were obtained for a progenitor model with $M_\mathrm{ZAMS}=16\,M_\odot$ (final mass $14.7\,M_\odot$) and an explosion energy of $E_\mathrm{ej}=10^{51}\,{\rm erg}$, and thus represent a single case.}
This result, in conjunction with the increasing number of observations indicating the presence of confined CSM around SNe II, suggests that dust formation in such environments may be a common and robust process. In addition, we calculated the dust emission light curves associated with our models and found them to be broadly consistent with observed mid-IR signals \modify{of SN~1998S}. These findings provide theoretical support for the CDS as a viable and potentially ubiquitous site of dust formation in interacting SNe. 
In what follows, we conclude by discussing several implications of our results and outline directions for future work aimed at refining dust formation models in confined CSM environments.

\subsection{Survival of Dust Grains against Shock Destruction}
\label{sec:dust_survival}
The survival of dust grains against sputtering caused by reverse shocks generated after the SN ejecta sweeps up the ISM has been a subject of extensive investigation \citep[e.g.,][]{Nozawa_et_al_2007,Silvia_et_al_2010,Slavin_et_al_2020}. In typical CCSNe, reverse shocks propagating into the ejecta heat and ionize the gas, leading to efficient thermal sputtering and potential destruction of newly formed dust grains. However, in the present scenario, dust forms in a CDS generated by the interaction between the SN ejecta and a confined CSM, which provides unique physical conditions.

One of the key features of the CDS in our models is its high density, which not only facilitates rapid dust formation but may also help dust grains grow to larger sizes before the arrival of the reverse shock. To illustrate this, we compare the characteristic density in the CDS with that in the inner ejecta, corresponding to the flat-density core in the broken power-law model described by Equation (\ref{eq:rho_ej}). 
The mean density of the inner ejecta, $\bar{\rho}_\mathrm{c}$, can be calculated by integrating Equation (\ref{eq:rho_ej}) from $r=0$ to $r=v_{t}t$,
\begin{eqnarray}
    \bar{\rho}_\mathrm{c}&=&\frac{3(n-3)M_\mathrm{ej}}{4\pi (n-\delta)(v_{t}t)^{3}} \nonumber \\
    &\sim&8\times10^{-16}\,{\rm g\>cm^{-3}}\left(\frac{M_\mathrm{ej}}{10\,M_\odot}\right) \nonumber \\
    &\times&\left(\frac{v_{t}}{4\times10^{8}\,{\rm cm\>s^{-1}}}\right)^{-3}\left(\frac{t}{500\,{\rm d}}\right)^{-3},
\end{eqnarray}
where $t\sim500\,{\rm d}$ is the time when the core has cooled down to $T\sim2000\,{\rm K}$ \citep[e.g.,][]{Nozawa_et_al_2003}.
Assuming that the entire core is composed of carbon, this density is comparable to the gas density of carbon presented in Figure \ref{fig:CDS_property} of Section \ref{sec:expected_dust_mass}, $X_\mathrm{C}\rho_\mathrm{CDS}\sim10^{-15}\,{\rm g\>cm^{-3}}$. However, we note that even higher densities can be realized in the CDS under certain conditions, particularly for models with shorter time from CSM ejection to CC $t_\mathrm{inj}$, which lead to an earlier onset of free expansion and thus allow the CDS to reach higher densities (see also Section \ref{sec:expected_dust_mass}).

Given that the dust grains formed in the CDS may survive reverse shock destruction, it is instructive to consider their potential contribution to the galactic dust reservoir over cosmic timescales. 
If each such event produces $\sim0.1\,M_\odot$ of dust, and if they occur at a rate of $\sim0.005\,{\rm yr^{-1}}$ in a galaxy \citep{Blanton_et_al_2003,Smith_et_al_2011,Graur_et_al_2017,Frohmaier_et_al_2021}, the cumulative contribution over $10\,{\rm Gyr}$ would reach $\sim5\times10^{6}\,M_\odot$. This corresponds to roughly $\sim5\%$ of the total dust mass observed in the Milky Way \citep{Draine_et_al_2007,Bovy_Rix_2013}, suggesting that such events may constitute a non-negligible, complementary source of galactic dust.

\modify{In addition, even if large grains form in the CDS, their survival is not guaranteed because reverse shocks can further process them into smaller grains that are more easily destroyed. \citet{Nozawa_Kozasa_2006} showed that SNe IIP can initially produce grains with radii of $\sim0.001$--$1\,{\rm \mu m}$. These grains are later exposed to reverse shocks as the supernova remnant evolves, and once the post-shock gas reaches temperatures above $\sim10^{6}\,{\rm K}$, thermal sputtering efficiently destroys small grain with radii of $\lesssim0.05\,{\rm \mu m}$. Shuttering can also fragment larger grains into smaller pieces. \citet{Nozawa_et_al_2007} further demonstrated that the survival probability of grains strongly depends on their size and their initial position within the ejecta, with small grains being much more vulnerable to destruction, especially in dense environments. Therefore, the growth of larger grains in the CDS is particularly important, as it enhances the chance that newly formed dust survives reverse shocks and ultimately contributes to the galactic dust reservoir.}

\subsection{Implications for Kilonova Surveys: Dust Emission as a Possible Contaminant}
\begin{figure}[t]
\centering
\includegraphics[width=\linewidth]{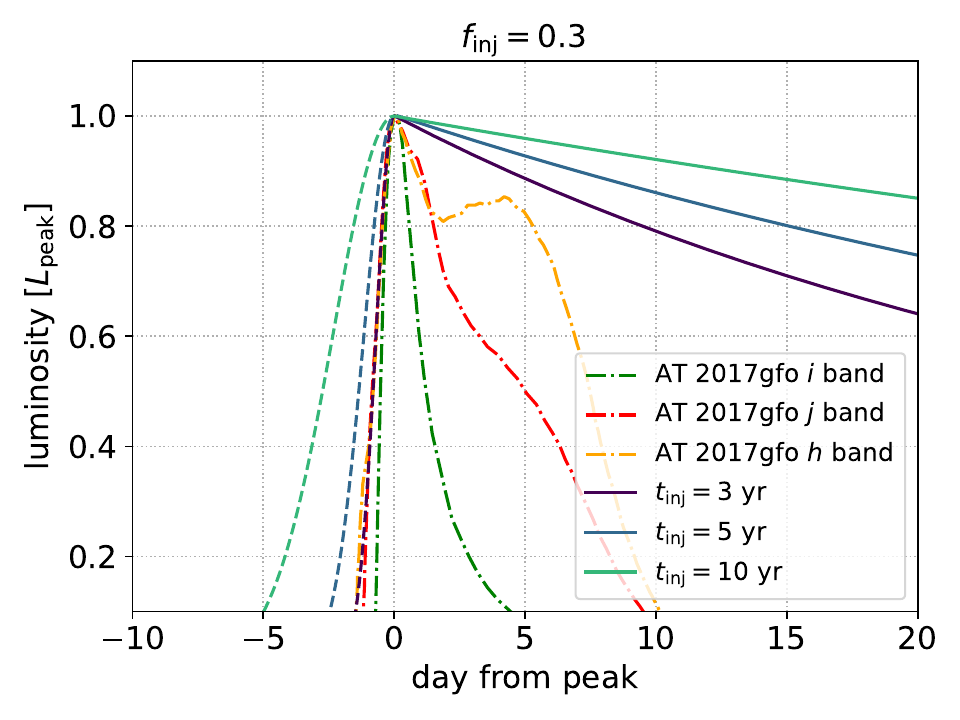}
    \caption{Comparison of the dust light curve with the $i,\,j,\,h$ band light curves of AT~2017gfo, scaled with the peak luminosity $L_\mathrm{peak}$. The rising phase, approximated by Equation (\ref{eqn:rise}), is indicated by dashed lines. The adopted parameter is $f_\mathrm{inj}=0.3$. The light curves of AT~2017gfo are taken from \citet{Kasen_et_al_2017}.}
\label{fig:cmp_KN}
\end{figure}
Recent studies have emphasized that several classes of fast optical transients (FOTs) can mimic the early-time behavior of KNe \citep[e.g.,][]{Bemmel_et_al_2025,Fulton_et_al_2025,Barna_et_al_2025}. For example, \citet{Fulton_et_al_2025} reported that some of luminous blue variable (LBV) outbursts exhibit rise times and decline rates comparable to those observed in AT~2017gfo, a KN associated with GW170817 \citep[e.g.,][]{Abbott_et_al_2017_AT2017gfo,Cowperthwaite_et_al_2017,Drout_et_al_2017,Valenti_et_al_2017}. In parallel, \citet{Barna_et_al_2025} performed a systematic analysis of FOTs, including a number of SNe IIb and other stripped-envelope SNe, and found that several events exhibit rise times of $\lesssim10\,{\rm d}$, comparable to those expected for KNe \citep{Kasen_et_al_2013}. Their results suggest that a diverse population of `impostor' transients may mimic the early photometric evolution of KNe in time-domain surveys. These findings reinforce the notion that rapid photometric rise alone is not an unambiguous signature of KNe.

In this context, our model introduces an additional source of potential photometric confusion in KN surveys. As shown in Section \ref{sec:IR_emission_from_dust}, once dust has formed within a CDS, the resulting IR emission can exhibit a steep rise over $\lesssim10\,{\rm d}$ due to the smaller radius of the confined CSM. 
This rising phase can be crudely approximated using the Gaussian function,
\begin{eqnarray}
    L_\mathrm{d}(t)\approx L_\mathrm{d}(\tau_\mathrm{cond})\exp\left[-\left(\frac{t-\tau_\mathrm{cond}}{\Delta \tau}\right)^{2}\right],\label{eqn:rise}
\end{eqnarray}
where $\Delta \tau=r_\mathrm{sh}(\tau_\mathrm{cond})/c$ is the characteristic timescale of the rising phase.
This rapid rise leads to light curves that are not easily distinguished from those of KNe, particularly in surveys lacking spectroscopic follow-up.
It should be noted, however, that these events can be distinguished by observing the decay phase due to the slower decay of the dust light curve compared to that of KNe, as we have shown in Figure \ref{fig:dust_luminosity}.
For comparison, we plot the representative light curves of dust scaled with the peak luminosity $L_\mathrm{peak}$ in Figure \ref{fig:cmp_KN}, together with $i\,(0.70$--0.85\,${\rm \mu m})$, $j\,(1.10$--1.35\,${\rm \mu m})$, and $h\,(1.50$--1.80\,${\rm \mu m})$ band light curve models of AT~2017gfo \citep{Kasen_et_al_2017}. 
While the decay phase is not well matched, this figure indicates that a comparable rising behavior can be obtained for specific parameter sets, i.e., shorter $t_\mathrm{inj}$.
\modify{It should also be noted that the assumption of instantaneous dust formation, introduced in Section \ref{sec:estimating_dust_luminosity}, may not hold exactly in reality. If the condensation process takes a finite timescale of a few days \citep[e.g.,][]{Nozawa_et_al_2003,Chiaki_et_al_2025}, the rising phase of the dust light curves could appear slightly more gradual.}

\subsection{Possible Caveats}
In this work, we assume that $T_\mathrm{d}$ is approximately equal to $T_\mathrm{gas}$ ($T_\mathrm{d}=(0.5$--$1)T_\mathrm{gas}$) throughout the evolution. This simplification neglects both radiative cooling of dust grains and collisional heating by gas particles. While this assumption may be justified in high-density regions where thermal coupling between gas and dust is strong, it may no longer be valid in optically thin or low-density environments where the energy exchange is inefficient. In reality, $T_\mathrm{d}$ is governed by the balance between these heating and cooling processes, as demonstrated in detailed models such as \citet{Nozawa_et_al_2008}. Given that the IR luminosity emitted by dust is highly sensitive to its temperature, the assumption of thermal equilibrium with the gas may introduce significant uncertainties in the predicted light curves under such conditions.

Another possible caveat is that we assume all the metals condense into dust, which may overestimate the total dust mass. In this work, we adopt solar metallicity for the composition of the ejecta of the hydrogen-rich envelope and CSM, under which oxygen is more abundant than carbon \citep{Asplund_et_al_2009}. Under such conditions, a certain amount of carbon can be locked up by the formation of the CO molecule, potentially suppressing the formation of amorphous carbon grains. Indeed, the observations of SNe II such as SN~1987A have shown that the formation of the CO molecule occurs $\sim100$--200$\,\mathrm{d}$ after the explosion \citep[e.g.,][]{Catchpole_et_al_1988, Spyromilio_et_al_1988,Bouchet_Danziger_1993}. In our model, where $X_\mathrm{C}\sim10^{-3}$, the maximum possible mass fraction of CO is estimated to be $(1+16/12)X_\mathrm{C}\sim2\times10^{-3}$, assuming the extreme case in which all available carbon is entirely locked up in CO molecules. In practice, however, the actual CO abundance may be lower due to the competition between formation and destruction processes \citep{Clayton_et_al_1999,Clayton_et_al_2001}, as well as the possibility that rapid cooling in the CDS allows dust condensation to occur before CO formation is complete. While our model does not solve the chemical reaction networks, it provides an upper limit on the dust mass under the assumption of complete condensation. Future work incorporating detailed chemical kinetics will be necessary to refine this estimate.

\subsection{Future Prospects}
As discussed in Section \ref{sec:dust_survival}, our results indicate that dust grains with relatively large sizes can form in dense CDS environments, but, as shown in previous studies, the subsequent interaction with reverse shocks generated after the SN ejecta sweeps up the CSM may significantly alter the initial size distribution. In particular, even large grains may be subject to fragmentation and erosion, leading to the formation of smaller grains that behave quite differently in terms of their dynamics.

A key challenge for future work is to determine the size distribution of dust grains based on physically motivated nucleation models and to follow how this distribution evolves through interactions with shocks. In this work, we assumed a fixed MRN distribution (Equation \ref{eqn:MRN}) as an approximation, but the actual grain sizes are expected to depend sensitively on the local thermodynamic conditions at formation and can subsequently change through post-formation processing.

This evolution of the size distribution has important implications for dust dynamics. \citet{Ragot_2002} showed that electromagnetic drag, arising from the emission of plasma waves by charged grains, becomes effective primarily for small grains with radii below $\sim0.01$--$0.1\,{\rm \mu m}$, depending on the ambient plasma conditions \citep[see also][]{Hoang_Lazarian_2012}. Large grains formed at early times are not significantly affected by this mechanism, but those that are later fragmented into smaller sizes may enter the effective range where electromagnetic slowing-down becomes important. To assess whether dust formed in confined CSM environments can be decelerated and retained within the host galaxy, future models should incorporate the coupling between grain size evolution and plasma drag. Such modeling would benefit from combining nucleation theory, shock-driven grain processing, and electromagnetic grain dynamics in a unified framework.

\begin{acknowledgments}
We thank the anonymous referee for valuable comments that greatly improved the manuscript.
We are greatful to Wataru Ishizaki for fruitful discussions, and Nozomu Tominaga for useful comments.
This work is supported by JSPS KAKENHI grant Nos.~22H00130, 23H04900, 23H05430, 23H01172.
\end{acknowledgments}

\begin{contribution}
YT was responsible for writing and submitting the manuscript, came up with the initial research concept and edited the manuscript.
KI and MS obtained the funding and edited the manuscript.
All authors participated in discussions.
\end{contribution}

\software{\texttt{MESA} \citep{Paxton_11,Paxton_13,Paxton_15,Paxton_18,Paxton_19,Jermyn_23}, \texttt{CHIPS} \citep{Takei_et_al_2022, Takei_et_al_2024}, Python libraries: Matplotlib \citep[v3.4.2;][]{Hunter:2007}, Numpy \citep[v1.17.4;][]{harris2020array}, Scipy \citep[v1.11.1;][]{2020SciPy-NMeth}}

\bibliography{dust}{}
\bibliographystyle{aasjournalv7}

\end{document}